\documentclass[nologo,a4paper,10pt]{ETHpaper}
\def\multiset#1#2{\ensuremath{\left(\kern-.4em\left(\genfrac{}{}{0pt}{}{#1}{#2}\right)\kern-.4em\right)}}

\title{Disentangling the Timescales of a Complex System: A Bayesian Approach to Temporal Network Analysis}

\renewcommand*{\thefootnote}{\fnsymbol{footnote}}
\author{Giona Casiraghi\footnote{Corresponding author, \texttt{gcasiraghi@ethz.ch}}, Georges Andres}
\authoralternative{Giona Casiraghi, Georges Andres}

\address{Chair of Systems Design, \\ ETH Zurich, Weinbergstrasse 56/58, 8092 Zurich, Switzerland \\[2mm] }
\www{\url{http://www.sg.ethz.ch}}

\newcommand{\G}[1]{\mathcal{G}_{#1}}
\newcommand{\tin}[1]{\theta^\text{in}_{#1}}
\newcommand{\tout}[1]{\theta^\text{out}_{#1}}
\newcommand{\tstar}[1]{\theta^\star_{#1}}

\newcommand{\kout}[1]{k^\text{out}_{#1}}
\newcommand{\kin}[1]{k^\text{in}_{#1}}

\begin{document}
\maketitle
\renewcommand*{\thefootnote}{\arabic{footnote}}

\begin{abstract}
Changes in the timescales at which complex systems evolve are essential to predicting critical transitions and catastrophic failures.
Disentangling the timescales of the dynamics governing complex systems remains a key challenge.
With this study, we introduce an integrated Bayesian framework based on temporal network models to address this challenge.
We focus on two methodologies: 
change point detection for identifying shifts in system dynamics, 
and a spectrum analysis for inferring the distribution of timescales.
Applied to synthetic and empirical datasets, these methologies robustly identify critical transitions and comprehensively map the dominant and subsidiaries timescales in complex systems.
This dual approach offers a powerful tool for analyzing temporal networks, significantly enhancing our understanding of dynamic behaviors in complex systems.

\textbf{Keywords:} Temporal Network Analysis, Bayesian Modelling, Change-Point Detection, Timescales, Complex Adaptive Systems, Resilience
\end{abstract}

\section{Introduction}

\begin{figure}[t]\centering
	\includegraphics[width=\textwidth]{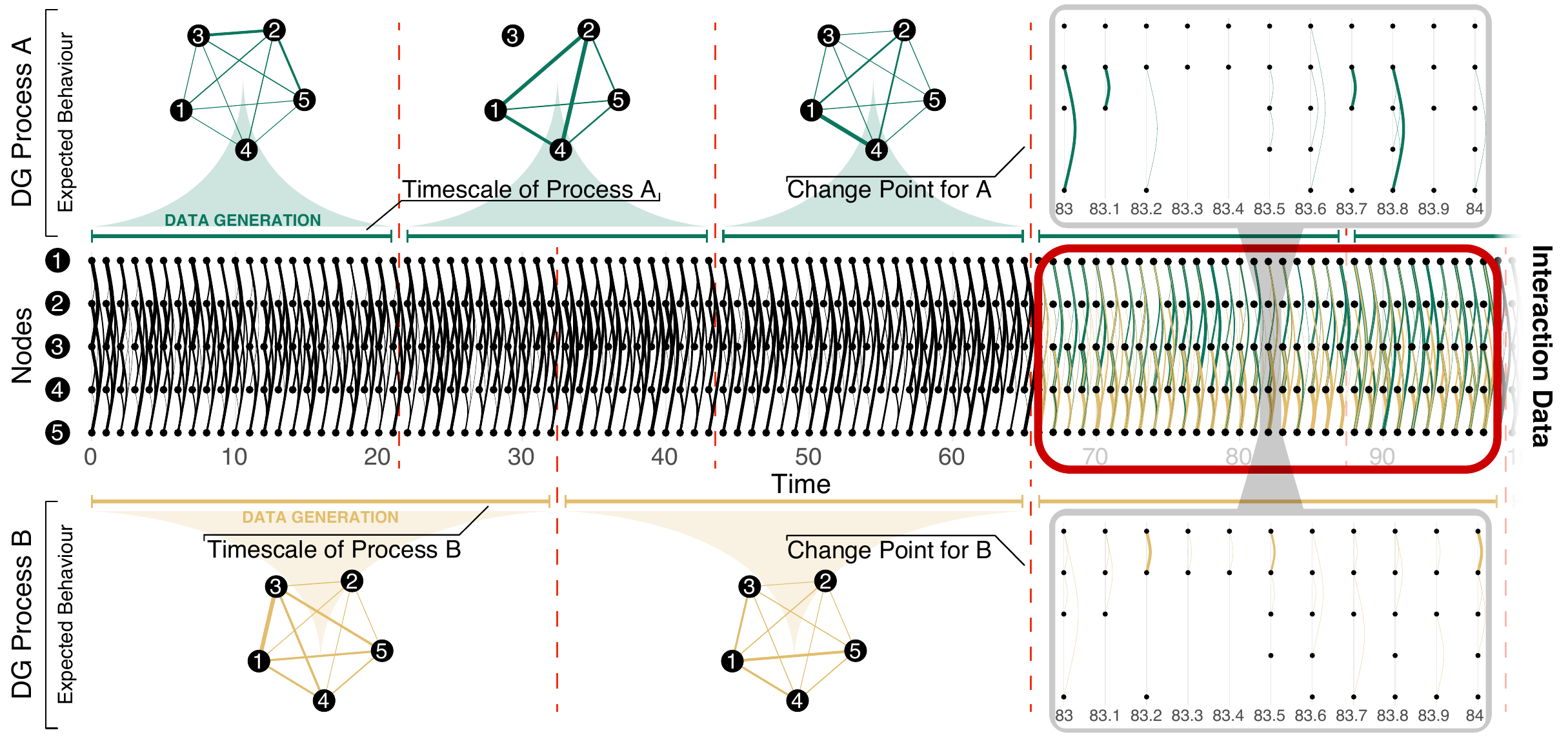}
  \caption{\textbf{Can we infer and disentangle timescales?}
  Interaction data originates from the complex interplay of multiple processes, highlighted in (green) and (yellow) in the figure.
  On the right side of the figure, we visualise how events originating from two distinct processes merge into the time-unfolded temporal network visualised in the centre.
  As the data-generating processes evolve at different timescales, we show the interactions expected from each process at any point in time as static networks (top and bottom rows).
  For the observer, edges coming from process A (green) or process B (yellow) are indistinguishable.
  However, the information about \emph{change points} (dashed red lines in the plot), i.e., points where data generating processes change, remains hidden in the data.
  In this article, we show how we can disentangle the different timescales at which complex systems evolve by means of Bayesian network models.
  }\label{fig:problem}
\end{figure}

Disentangling and understanding the timescales of the concurrent processes underlying a complex system is crucial for anticipating critical transitions and managing disruptions, be it in ecosystems~\cite{scheffer2003catastrophic}, social~\cite{scheffer2021loss}, or climate systems~\cite{dakos2008slowing}.
Studies have shown that variations in a system's dynamics can serve as an early-warning signal for impending catastrophic events~\cite{scheffer2009early} or as a necessary adaptation to recover from large disruptions~\cite{ambra2024}.
These results highlight the necessity for advanced methodologies capable of disentangling the timescales of the different processes underlying complex systems.

Key systemic properties like resilience, defined as a system's capacity to withstand and respond to disturbances~\cite{resilience_mega_paper}, hinge on the complex interplay of different processes operating across a spectrum of timescales~\cite{scheffer2009early,falk2019scaling}.
These processes span from immediate, transient events to slow, long-term evolutions~\cite{bar2019dynamics}.
Building on this understanding, we illustrate in \cref{fig:problem} a system driven by two distinct processes, each operating at its own timescale.
The evolution of these processes is depicted as a series of networks, each capturing the interaction patterns among the system's elements.
The actual interaction data--visualized as a time-unfolded network in the middle panel--stem from the interplay of both processes.
For the observer, however, interactions originating from the two processes are indistinguishable.
This poses a significant hurdle to the analysis of the system, as each process evolves at a distinct timescale, one being considerably slower than the other.
Such an interplay masks the true nature of interactions, making it hard to discern the impact of the different processes on the overall system dynamics.
Identifying and separating these processes is crucial for a comprehensive analysis of the system.

To achieve this, we need to disentangle the timescales of these concurrent processes.
How can we do so without prior knowledge about the processes generating the observed interaction data?
Temporal networks are a crucial tool addressing this challenge~\cite{scholtes2014causality}.
To model system evolution, temporal networks map interactions to sequences of events \((v, w, t)\)~\cite{holme2019temporal}, providing deep insights into temporal dynamics~\cite{Holme2012,masuda2017random}.
While existing methods excel in tasks like community detection~\cite{rosvall2008maps,delvenne2010stability,de2015identifying,stanley2016clustering} or epidemic spreading~\cite{Peixoto2018change,Valdano2015,Masuda2016}, they often fall short in capturing the full spectrum of timescales that govern complex systems.

Current approaches in temporal network analysis, spanning dynamic community detection to multiscale and multilayer network modeling, have made significant strides in understanding the structure and dynamics of temporal networks~\cite{bovet2022flow, mucha2010community, de2015identifying, matias2017statistical, gauvin2014detecting, ghasemian2016detectability, xu2013dynamic}.
A common thread among these methods is their focus on identifying change points within the network~\cite{deRidder2016detection,peixoto2017modelling,peixoto2015inferring,Peixoto2018change}.
These change points are crucial for understanding shifts in network dynamics, reflecting moments when the underlying system's behavior undergoes significant changes~\cite{corneli2018multiple,peel2015detecting}.
However, while adept at pinpointing these critical points in time, these approaches do not investigate further the dynamics of the network structure itself.
They tend to focus on the 'what' and 'when' of changes but less on `how' these changes unfold over different timescales.
This results in a gap in comprehensively understanding the often overlapping processes that drive the evolution of the network over time. 

With this work, we propose a change of perspective in the analysis of temporal networks.
Building upon the methodological advances proposed by \citet{peixoto2015inferring}, we introduce a Bayesian approach that goes beyond the mere identification of change-points.
While the latter remains a crucial step in any temporal network analysis,
our method employs temporal configuration models to further infer the \emph{spectrum of timescales} that drive the network's dynamics.
With such an approach, we can unravel the processes occurring at different timescales within a system, providing a more profound understanding of its inherent dynamics.

Our results demonstrate the effectiveness of this methodology in capturing the complex interplay of timescales.
We illustrate this through the analyses of various datasets.
For example, in analysing the evolution of email communication within the Enron Corporation, we observe a distinct shift in dominant timescales during the unfolding of the Enron scandal, providing a quantitative measure of the organizational disruption during this period.
Similarly, in a transversal analysis of 101 co-editing networks of large Open Source Software projects, we find that a slowdown in network dynamics precedes the departure of core developers.
When projects successfully adapt and recover from this shock, their dynamics revert to a pre-shock condition.
In contrast, projects that fail to recover from the shock continue to experience a slowdown until they ultimately fail.

This work not only deepens our understanding of temporal networks but also opens new avenues for exploring the dynamic nature of complex systems, as we uncover their underlying spectrum of timescales.

\section{Our Framework}
\subsection{A General Bayesian Model}

The core challenge of this work is to effectively disentangle the various timescales that shape the dynamics of a temporal network.
To tackle this, we employ a framework that simultaneously models the interactions observed within a temporal network $\mathcal{G}_T$ and the evolution of the processes generating these interactions.
This dual focus is expressed through a joint probability distribution, which serves as the cornerstone of our approach:
\begin{equation}\label{eq:jointgeneral}
\Pr(\mathcal{G}_T, \bm{\gamma}, \bm\Delta) = \Pr(\mathcal{G}_T|\bm{\gamma}, \bm\Delta) \cdot \Pr(\bm{\gamma}| \bm\Delta) \cdot \Pr(\bm\Delta)\;.
\end{equation}
Each component of this equation models a distinct aspect of the temporal network:

\begin{enumerate}
    \item \textbf{Likelihood of Observable Interactions $(\Pr(\mathcal{G}_T|\bm{\gamma}, \bm\Delta))$}: This term quantifies the probability of observing the specific interactions recorded in the network, $\mathcal{G}_T$, given a set of parameters $\bm{\gamma}$ and timescales $\bm\Delta$. It essentially measures the fit of our model (defined below) to the observed interaction within the temporal network.
    
    \item \textbf{Prior Probability of Model Parameters $(\Pr(\bm{\gamma}| \bm\Delta))$}: This component represents our prior knowledge or assumptions about the parameters $\bm{\gamma}$, conditioned on the timescales $\bm\Delta$. It integrates our theoretical understanding or previous findings about the data-generating processes into the model.
    
    \item \textbf{Prior Probability of Timescales $(\Pr(\bm\Delta))$}: The final term accounts for our initial hypotheses about the potential timescales $\bm\Delta$, which influence the network's dynamics. This term is usually set as a uniform distribution, reflecting a non-biased stance towards any particular timescale before data analysis.
\end{enumerate}

\subsection{Expressing the Joint Probability Distribution}

We need to define the individual components of the joint probability distribution outlined in \cref{eq:jointgeneral}.
To do so, our framework exploits a common assumption in temporal network analysis~\cite{holme2019temporal}.
Namely, that observable interaction events occur at a faster rate compared to the evolution of the underlying data-generating processes.
This distinction allows us to model the temporal network as a sequence of static networks, effectively aggregating the ``fast dynamics'' without losing sight of the overarching ``slow dynamics'' that drive the system's evolution.

Following this assumption, we introduce a time-window partition \(\bm\Delta\), dividing the entire observation period $[0,T]$ into $z$ adjacent intervals.
Within each interval, the interaction data is aggregated into a static network, ignoring the temporal ordering of events.
This modeling choice leads to a series of aggregated sub-graphs $\{\G{\tau}\}_{\tau\in [1,z]}$, each representative of the network within its respective time-window.

Assuming independence between these time-windows, the temporal network $\G T$ can be modeled as:
\begin{equation}\label{eq:tmodel}
  \Pr(\G T|\pmb\theta,\pmb\Delta) = \prod_{\tau=1}^z \Pr(\G \tau|\pmb\theta_\tau)\,.
\end{equation}
where $\Pr(\G\tau|\pmb\theta_\tau)$ is defined by a static network model parametrized by $\pmb\theta_\tau$ for a given time-window $\tau$ that we define below.

\paragraph{Temporal Configuration Models}

To express $\Pr(\G\tau|\pmb\theta_\tau)$, we look for the simplest model that allows us to capture significant changes in the structure of the temporal network.
We argue that the solution lies in monitoring the activity patterns of nodes in the network.
These activity patterns, whether stemming from global shifts like seasonal variations or local dynamics like community formation, will be reflected in the sequences of node degrees.

Therefore, we focus on \emph{configuration models}~\cite{fosdick2018}.
Configuration models are simple and efficient models adept at preserving the degree sequences observed within our defined time-windows.
Here, we focus on the Hypergeometric Configuration Model (HCM).
Adapting this framework to other configuration models is straightforward.

The HCM is formulated as follows~\cite{Casiraghi2021}:
\begin{equation}\label{eq:hcmmain}
\Pr(\mathcal{G}) = {\binom{\sum_{vw}\tout{v}\tin{w}}{m}}^{-1} \prod_{vw}{\binom{\tout{v}\tin{w}}{A_{vw}}}\;,
\end{equation}
where $\tout{v}\tin{w}\in\mathbb N_0$ represents possible edges between nodes $v$ and $w$, with $\tout v$ and $\tin v$ indicating each node's respective outgoing and incoming activity levels (see \cref{sec:methods:htcm} for more details).

By incorporating the HCM into \cref{eq:tmodel} (see \cref{sec:methods:htcm}) and specifying uniform priors on the parameters, we arrive at an explicit definition of the joint probability \cref{eq:jointgeneral} as:
\begin{equation}\label{eq:htcm_joint}
  \Pr_\text{HTCM}(\mathcal{G}_T, \bm{\theta_{\tau}}, \bm\Delta) \propto
  \Pr(\bm\Delta)
  \prod_{\tau=1}^{z}
  \frac{m_\tau!}{\binom{m^2_\tau}{m_\tau}}
  \left(\frac{(N-1)!}{(N+m_\tau-1)!}\right)^{2}
  \prod_{vw}\frac{(\tout{v\tau}\tin{w\tau})!}{(\tout{v\tau}\tin{w\tau}-A_{vw\tau})!}\;.
\end{equation}
A full derivation of the joint probability distribution is detailed in \cref{sec:methods:bayesian}.
In the following sections, we highlight how to choose $\Pr(\bm\Delta)$ depending on whether change-point (\cref{sec:cpd}) or timescale detection (\cref{sec:tsd}) is of interest.

\section{Change-point detection}\label{sec:cpd}
\subsection{The Minimum Description Length Principle}

The ability of the model presented in \cref{eq:htcm_joint} to accurately represent $\G T$ largely depends on the alignment of the \emph{transitions between time-windows} with the \emph{changes in the data-generating processes}.
These changes are usually referred to as change-points: critical moments when the data-generating processes undergo significant transitions~\cite{peel2015detecting}.
Consider the example in \cref{fig:problem}, where two data-generating processes change at distinct intervals: one every 22 time-steps (green) and another every 33 time-steps (yellow).
By defining the time-window partition $\bf\Delta$ such that our model updates its parameters every $22,11,22,\dots$ time-steps, we can precisely pinpoint all \emph{change-points} in the network dynamics.
Thereby, the model captures all critical shifts of the underlying data-generating processes.

The main challenge lies in inferring an optimal time-window partition $\bf\Delta$ without explicit knowledge of the data-generating processes.
$\bf\Delta$ must be detailed enough to capture the processes' change-points, yet not so much as to overfit the data.
Striking this balance is key to ensuring that the model accurately reflects the underlying temporal dynamics ignoring noise.

This balance can be achieved by employing the concept of description length (DL)~\cite{rissanen1983}.
The DL quantifies the amount of information necessary to represent a dataset by means of a given model.
By balancing model granularity and explanatory power, the DL serves as a gauge for model quality.
The DL, denoted as $\Sigma$, is defined as the negative logarithm of the joint probability of the data $\mathcal{G}_T$ and the model parameters $\bm{\gamma}$:
\begin{equation}
\Sigma = - \log_2 \Pr(\G T,\bm{\gamma})\;.
\end{equation}
A model that minimizes $\Sigma$ is optimal, as it effectively captures the dynamics of the dataset without overfitting or underfitting the data.

In the model delineated in \cref{eq:htcm_joint}, the complexity and the potential to identify change-points are governed by the time-window partition $\bm{\Delta}$.
A model with more time-windows has a higher chance of identifying the data-generating process' change-points, but it also runs the risk of overfitting.
To determine a balanced partition, we adopt a uniform prior over all time-window partitions \(\bf\Delta\), which is expressed as: 
\begin{equation}\label{eq:priordelta}
\Pr(\bm\Delta) = {\binom{T-1}{z-1}}^{-1}\frac1{T}\,.
\end{equation}
This formula represents a uniform distribution across all potential partitions for any number of time-windows $z$, ranging from 0 to $T$.
The derivation, inspired by the 'stars and bars' problem in probability theory~\cite{pitman2012probability}, is detailed in \cref{sec:methods:bayesian}.

With this general prior, the model that minimises the description length yields an optimal time-window partition $\bm{\hat\Delta}$.
Under ideal conditions, $\bm{\hat\Delta}$ aligns with all relevant change-points. 
This is exemplified in an application of the method to our case study \cref{fig:problem}, where the $\bm{\hat\Delta} = (22,11,22,11,\dots)$ perfectly aligns with the change-points of the underlying processes.
To identify such $\bm{\hat\Delta}$, we implement a Markov-Chain-Monte-Carlo (MCMC) algorithm.
Comprehensive details and methodology of this algorithm are elaborated in \cref{sec:mcmc}.

\subsection{Application to data}\label{sec:cpd:data}
\begin{figure}[ht]
  \centering
  \includegraphics[width = .45\textwidth]{ 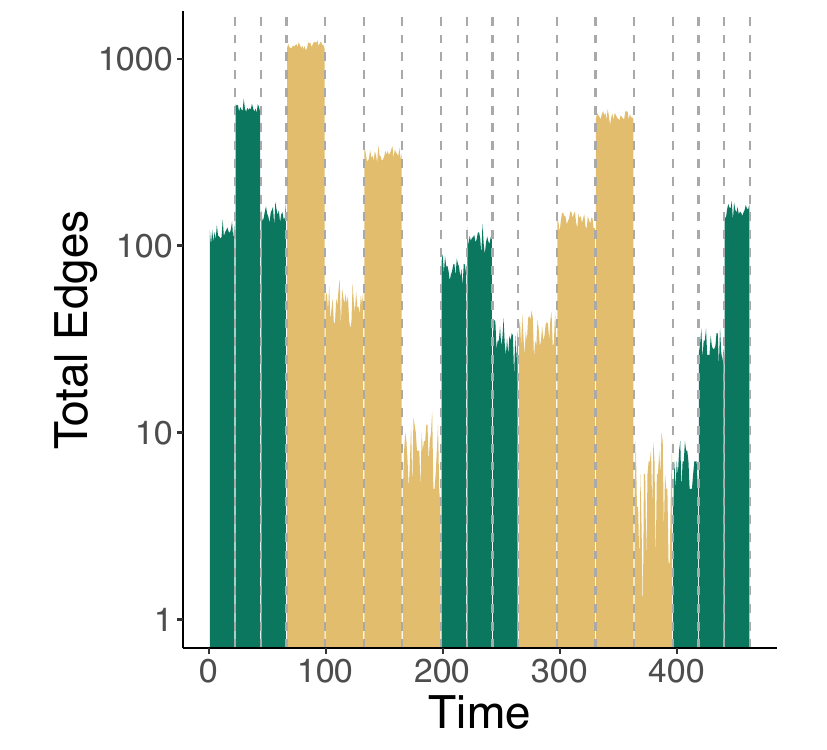}\hfill
    \includegraphics[width = .45\textwidth]{ 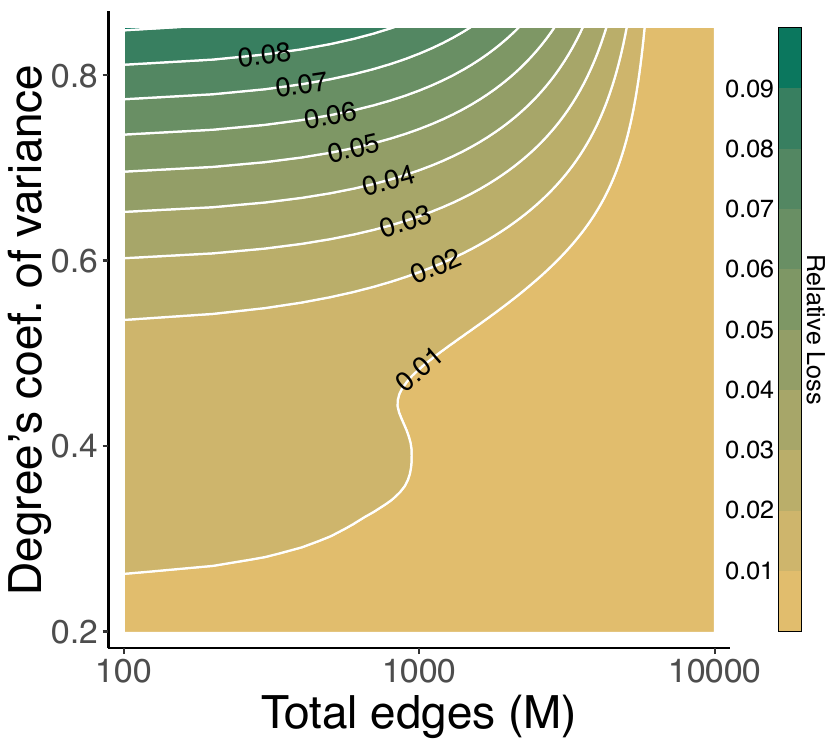}
  \caption{Investigating the resolution limits of change-point detection. 
  \textbf{(Left)} 
  Inferred time-windows from synthetic data with ground truth $\Delta_{GT} = (22,22,22,33,33,33,33,22,...)$.
  Different colors are assigned to the two different timescales $\psi_1=22$ and $\psi_2=33$ used to generate the data.
  The inferred partition is shown as a dashed line.
  When the number of interactions $M$ is large enough (here $M = 100\,000$, $N=20$, $T=462$), we find the correct partition.
  \textbf{(Right)} 
  This contour plot illustrates the relative loss in bits between the ground truth partition and the optimal partition found by our change-point detection method.
  We vary the number of interactions ($M$) and the heterogeneity of the data, represented by the coefficient of variance of degrees.
  Green regions denote high loss, transitioning to yellow for lower loss. 
  Adequate data volume ($M$ large) ensures precise model performance, accurately identifying the true decomposition.
  However, as the data volume diminishes and degree variance increases, the model's accuracy in detecting the correct partitions decreases, leading to more pronounced errors.
  This analysis is conducted with $N = 20$ and $T=105$.
  }\label{fig:mcmc synthetic}
\end{figure}

\paragraph{Synthetic Data}
To test the limits of our method's resolution, we create a synthetic case-study featuring two distinct, non-overlapping timescales: a shorter one at $\psi_1$ time-steps and a longer one at $\psi_2$ time-steps.
This setup simulates a scenario where the interaction pattern changes periodically.
The details are provided in \cref{sec:methods:data:synth}.
Our ground-truth partition is $\Delta_{GT} = (\underbrace{\psi_1,\psi_1,\psi_1},\underbrace{\psi_2,\psi_2,\psi_2,\psi_2},\underbrace{\psi_1,...})$.
If enough interactions are recorded, our method easily identifies this pattern and accurately mirrors the ground truth in its partitioning, as illustrated in \cref{fig:mcmc synthetic} (left).
However, as the number of recorded interactions decreases and the heterogeneity in the data increases, accurately detecting the change-points becomes harder.
The contour plot in \cref{fig:mcmc synthetic} (right) visualizes this. 

\paragraph{Empirical Data -- ENRON}
The ENRON dataset~\cite{zhou2007strategies,perry2013point}, comprises corporate email communications exchanged within the American energy company Enron Corporation.
Enron, once considered a blue-chip stock and a highly innovative company, was involved in accounting fraud and corruption.
As news of widespread fraud became public throughout 2001, the company filed for bankruptcy on December 2, 2001.

\begin{figure}[ht]
  \centering
    \includegraphics[width = .45\textwidth]{ 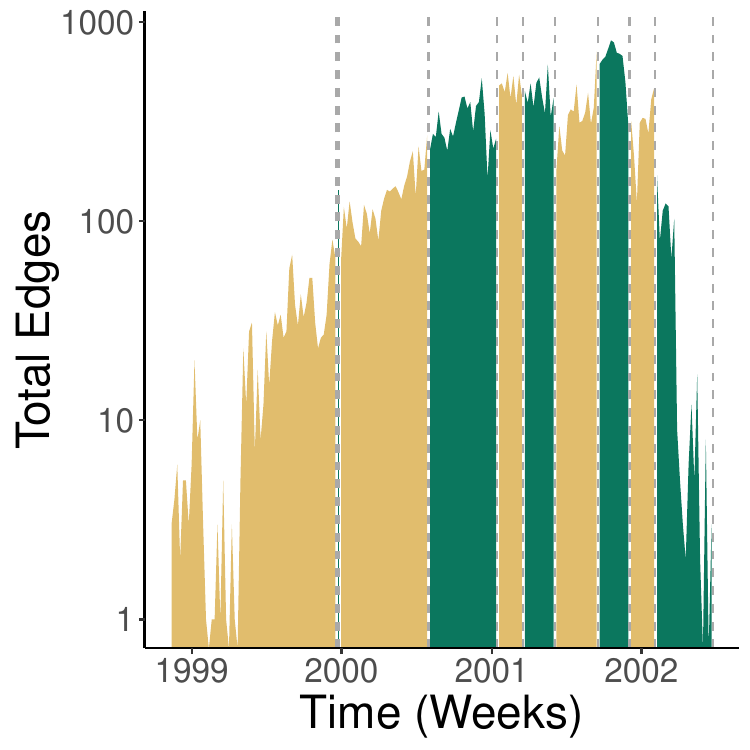}\hfill
    \includegraphics[width = .45\textwidth]{ 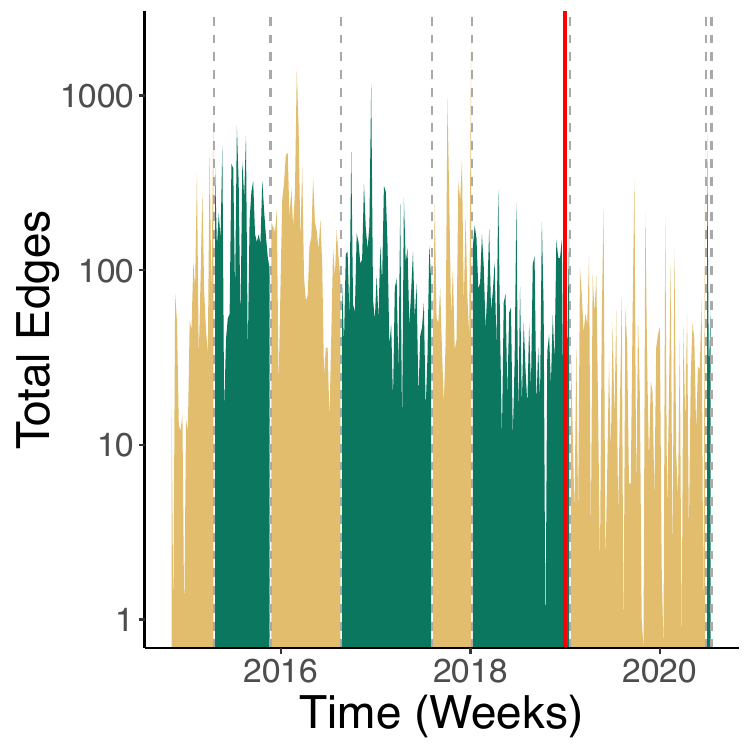}
    \caption{Optimal time-window partitions for ENRON \textbf{(Left)} and of the \texttt{fastlane} project in the DEVS dataset \textbf{(Right)}. 
    In ENRON, there is an evident acceleration in the communication dynamics from approximately 2001 onwards, marking the beginning of the accounting scandal.
    For DEVS, we observe a regular pattern in the time-window partition.
    The red line indicates when a core-developer suddenly stopped committing to the GitHub project.
    This event is identified as a change-point in the optimal partition that sets a transition in the project dynamics right after the shock.}\label{fig:mcmc data}
\end{figure}

The ENRON dataset serves as a real-world test bed for our method:
Initially, interaction patterns span longer time-windows, indicating a relatively stable communication structure, as inferred by our model and visualised in \cref{fig:mcmc data}~(left).
However, as the infamous ENRON scandal unfolds the communication dynamics accelerates.
This is reflected in the emergence of shorter time-windows in the optimal partition.
This temporal shift could be indicative of the organizational turmoil and heightened internal communications as the company grappled with the unfolding scandal.

\paragraph{Empirical Data -- DEVS}
The DEVS dataset tracks developer contributions to a selection of 101 large Open Source Software (OSS) project on GitHub, all characterized by a critical event: the sudden halting of contributions from a core developer~\cite{russo2024shock}, i.e., a developer holding a central role within the project.
This event is hypothesized to have a substantial impact on the project's dynamics~\cite{gote2023locating,russo2024shock}, and possibly even jeopardize its survival~\cite{oliveira2020code}.

This dataset offers a unique perspective on how significant events within a project can be captured by means of change-point detection methods.
In \cref{fig:mcmc data}~(right), we visualise the optimal time-window partition detected for the project \texttt{fastlane}.
Remarkably, our method identifies a significant change in the network's dynamics shortly after the departure of the core developer (red line).
This finding is not isolated to this project alone; similar patterns are observed in most of the 101 projects analysed.
More specifically, we identify a change-point at most one week after the departure of the core-developer in 63 out of the 101 dataset.
Further, in 90\% of the projects change-points can be found at most 2 months  after the departure of the core-developer.
This corresponds to a negligible amount of time (less than 3\%) compared to the average duration of the projects \emph{before} the core-developer departure (288.4 weeks on average, with 20\% of the project taking more than 400 weeks).

These results underscore the potential of change-point detection methods in identifying sudden shifts in the interaction dynamics driven by key individual contributors or events.

\section{Disentangling Timescales}\label{sec:tsd}

The application of the Minimum Description Length principle in change-point detection has proven effective in identifying transitions within both synthetic and empirical temporal networks.
This efficacy is particularly evident when aligning the model's outcomes with established ground truths and key events, as demonstrated in the analysis of datasets like ENRON and DEVS.
However, this method falls short in disentangling the distinct timescales of the underlying processes driving these changes.
This gap underscores the necessity for a more comprehensive approach.

\subsection{The Timescales Spectrum of a Temporal Network}
Drawing parallels to signal processing, we define the \emph{spectrum} of a temporal network's dynamics as the range of timescales that shape the dynamics of the data-generating processes underlying the network.

A timescale may be defined as the average time at which a data-generating process changes~\cite{holme2019temporal}.
As a shift in the data-generating process manifests as a change-point, we can encode timescales as regular time-window partitions.
Specifically, we restrict the model to partitions of time-windows of equal size $\Delta$, where $\Delta$ is the timescale that we want to encode.
Therefore, to evaluate a timescale of length $\Delta$, we define the partition $\bf\Delta$ as the sequence $(\alpha,\Delta,\Delta,\Delta,\dots,\Delta,\omega)$.
Because the starting point of the data $t=0$ may not align with a change-point in the data-generating process, we require a padding $\alpha$ to allow our model to optimally fit the data.
Similarly, the last window may be of size $\omega\leq\Delta$.
The uniform prior on the set of these possible partitions is given by
\begin{equation}\label{eq:priordelta_fixedTW}
  \Pr(\bm\Delta) = \begin{cases}
 	\left(\sum_{x = 1}^{T-1} x\right)^{-1} = \left(\frac{T(T-1)}{2}\right)^{-1} \qquad&\text{if $\Delta_i = \Delta_j = \Delta$ $\forall i,j\in[2,z_\Delta]$}\\
 	0 \qquad&\text{otherwise.}
 \end{cases}
\end{equation}

The partition $\hat\Delta$ that yields the minimum description length is considered optimal.
It most effectively averages the overall temporal evolution of all data generating processes involved.
In an ideal scenario where the partition given by \(\hat\Delta\) \emph{exactly matches} all change points, \(\hat\Delta\) has to be the greatest common divisor (GCD) among the timescales of all data-generating processes.
Of course, in this ideal case, \emph{all} common divisors will match all change-points.
However, only the GCD provides the minimal number of time-windows to do so, yielding the global minimum of the DL function.

Furthermore, since the space of all partitions scales with the square of $T$, we can efficiently compute the description length for \emph{every} partition defined by $\Delta \in [1,T]$.
Following the reasoning detailed in \cref{sec:methods:thms}, we find that every partition aligning with some change-points will appear as a local minimum of the DL function.
Additionally, local minima of the DL function will correspond to multiples or divisors of the optimal partition \(\hat\Delta\).
Therefore, the partitions corresponding to the timescales of each data-generating process will appear as a local minimum of the DL.
The remaining local minima will further correspond to the harmonics of these true timescales and to their resonances, i.e., to combinations of these that match groups of change points.

We argue that the set of all partitions appearing as local minima of the DL defines the \emph{spectrum} of \(\G{T}\)'s dynamics.
Similarly to the spectrum of a signal, the spectrum of \(\G{T}\) contains all timescales needed to describe the dynamics of the temporal network.
Clearly, the timescales of the data-generating processes, \(\hat\Delta\), their harmonics, and their resonances are not equally relevant for understanding these dynamics.
As our aim is to disentangle the timescales of the data-generating processes from the others, we must assign an importance score to each of the timescales in the spectrum of $\G T$.
Borrowing again from signal processing, we employ the concept of \emph{topographic prominence}, ranking timescales according to their local importance in the DL landscape, as explained in \cref{sec:methods:thms}.
The concept of peak prominence enables us to discern influential timescales from those that are less critical or simple artifacts.
We can then represent the spectrum of $\G T$'s dynamics in a \emph{spectrogram}, as shown in \cref{fig:spectrogram}.

\subsection{Application to Data}\label{sec:spectrum:data}
\paragraph{Synthetica Data}

\begin{figure}[t]
\includegraphics[width=\textwidth]{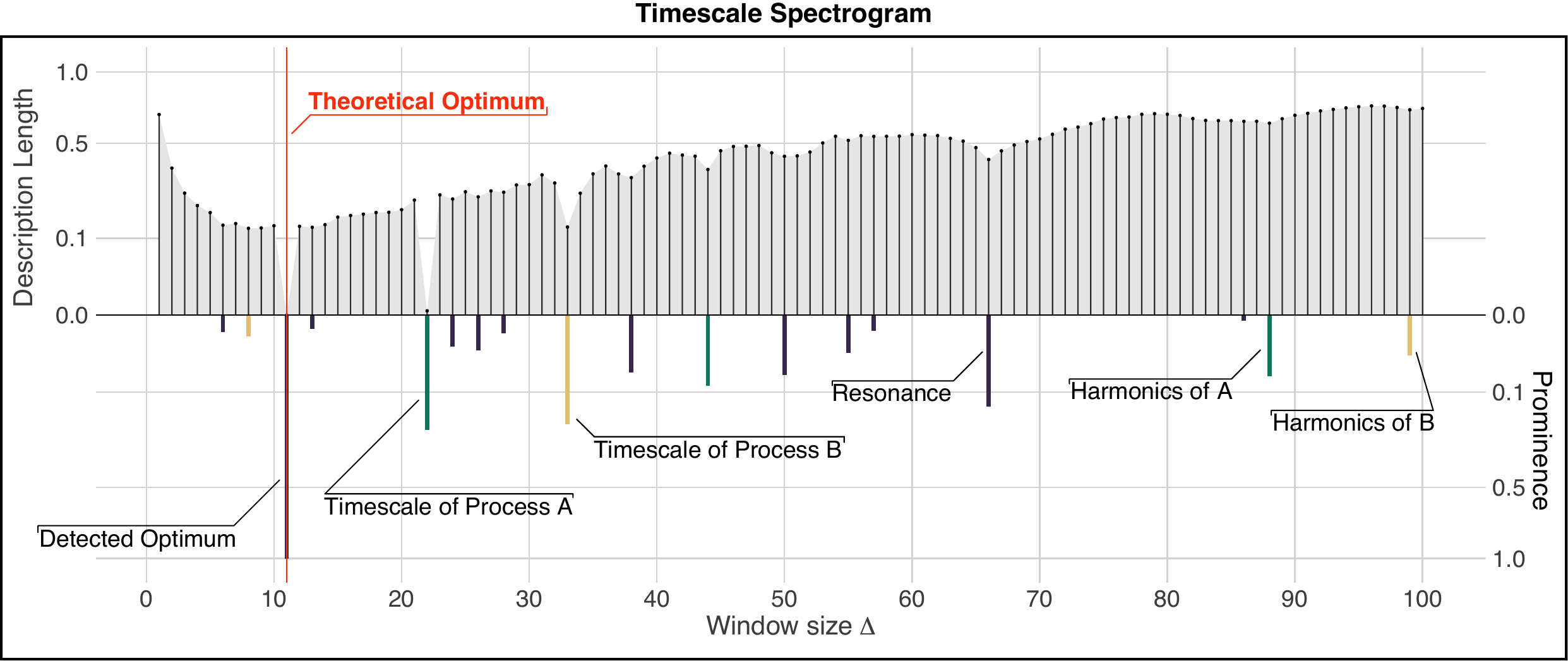}
  \caption{\textbf{Timescale Spectrogram.}
  The process described in \cref{fig:problem} leaves us with a temporal network $\G{T}$ recording the interactions generated by processes A (green) and B (yellow).
  We partition $\G{T}$ into adjacent time-windows of size $\Delta$.
  By plotting the \emph{description length} of a temporal network model as a function of the time-window size $\Delta$, we obtain the \textbf{(top)} plot.
  The minimum description length (normalised to 0 in the plot) gives us the optimal time-window partitioning for the data.
  This value ($\Delta=11$) matches with the theoretical optimum, the greatest common divisor between the timescales of process A ($\psi=22$) and B ($\psi=33$).
  Further, we can observe different local minima in the description length.
  By plotting their topographical prominence \textbf{(bottom)}, we obtain a \emph{spectrogram} for the different timescales present in the temporal spectrum of the data.
  Similarly to a classical spectrogram, this plot allows us to identify the most important frequency in the temporal data that we study.
  The second most prominent minima match the two timescales of the data-generating processes.
  The remaining local minima correspond to harmonics and sub-harmonics of these fundamental frequencies, shown in (green) and (yellow), or combinations of them, shown in (gray).
  }\label{fig:spectrogram}
\end{figure}

With these new tools in hand, we revisit the running example of \cref{fig:problem}.
There, two processes are simultaneously generating interactions at different timescales.
The shorter timescale is of 22 time-steps, i.e. every 22 time-steps, we change the parameters of the data-generating process yielding different interaction patterns.
The longer timescale is of 33 time-steps and we overlay these generated interactions to the previous ones.

This process results in the temporal network $\G T$ visualized in the center panel of \cref{fig:problem}.
The green links were generated from the shorter timescale, the yellow from the longer timescale.
In \cref{fig:spectrogram}, we plot the DL as a function of every possible partition of window size $\Delta\in[1,100]$.
The DL profile has prominent troughs at three very specific time-window lengths: 11, 22, and 33.
The latter two are exactly the timescales of the data-generating processes.
Furthermore, the most optimal partition is at 11, the GCD of these two timescales.

On the bottom part of~\cref{fig:spectrogram} we show the results obtained by computing the prominence of the spectrum of $\G T$.
The three most prominent timescales are indeed the three aforementioned: 11, 22, and 33.
Further, the spectrogram allows us to understand what the additional timescales in the spectrum represent: they are resonances and harmonics of the true timescales.
We group the harmonics of each of the two timescales by color.
The less prominent values are to be attributed to either noise coming from the stochastic data-generating processes, or to finite-size effects.

\paragraph{Empirical Data -- ENRON}
\begin{figure}[ht]
  \centering
  \includegraphics[width = .49\textwidth]{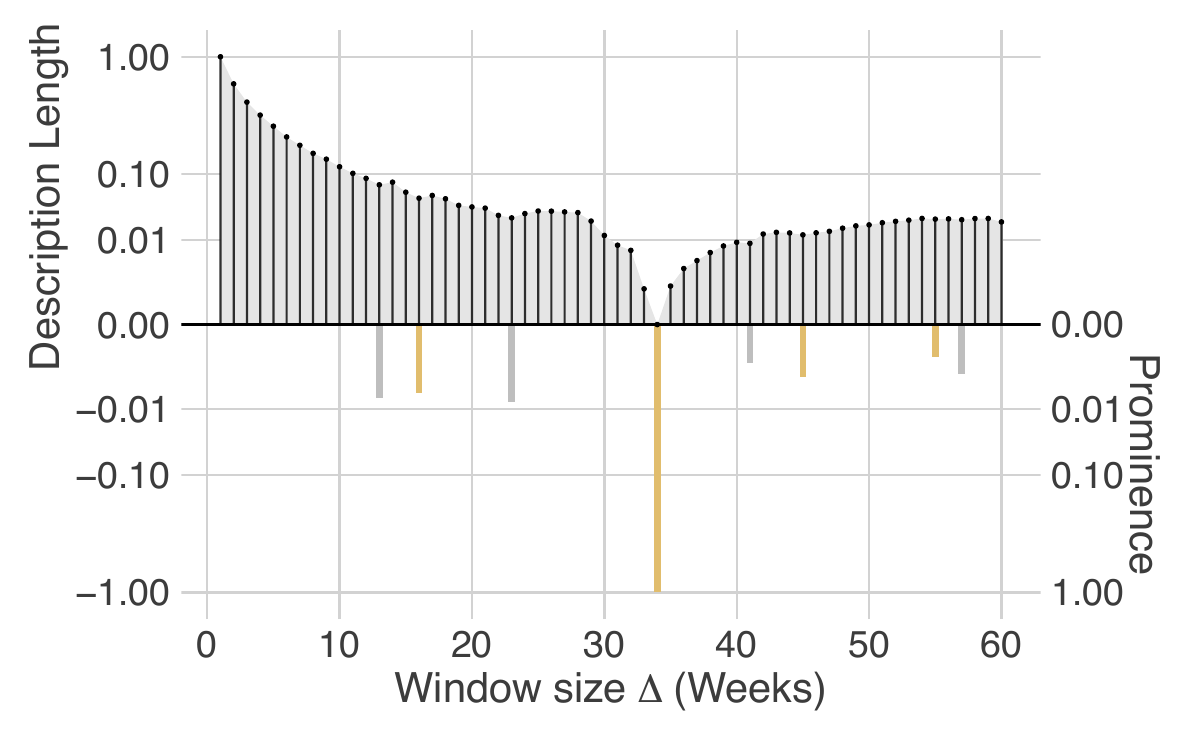}\hfill
  \includegraphics[width = .49\textwidth]{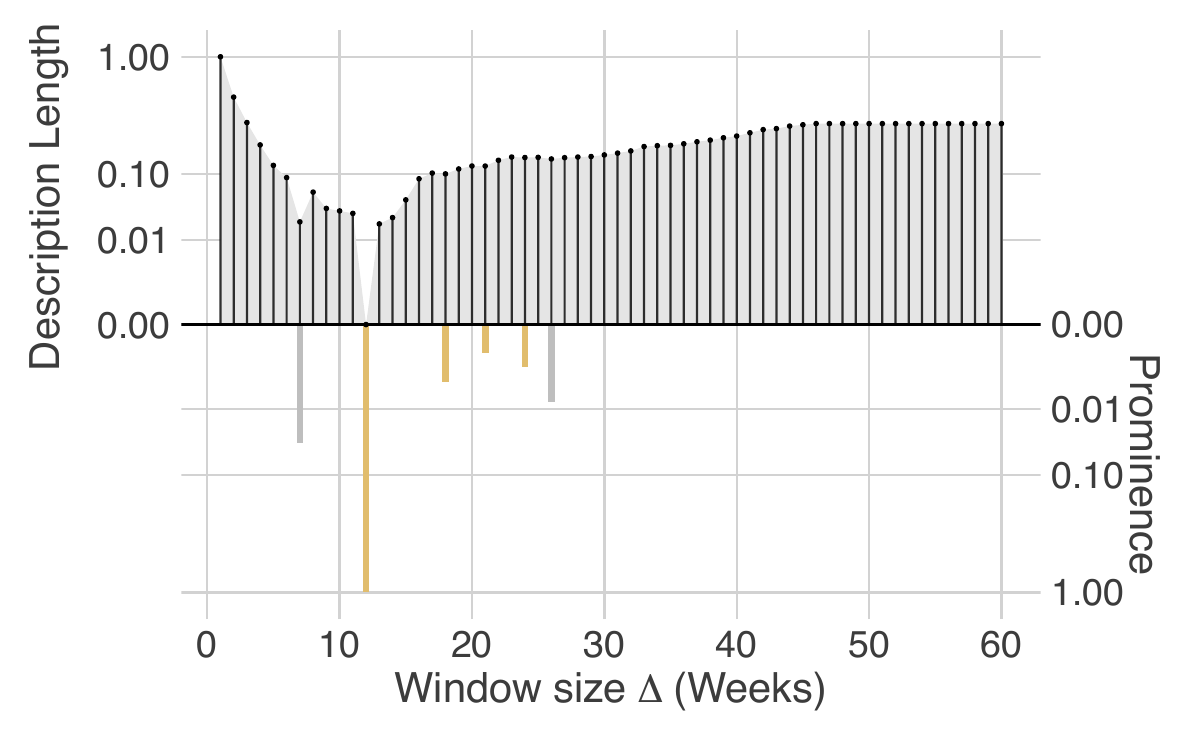}
  \caption{Spectrograms for the ENRON dataset computed on the communication network obtained on the data before 2001 (left) and after 2001 (right).
  In the top part of the plot, we visualise the normalised DL as a function of the fixed window size $\Delta$ that defines the timescale analysed.
  In the bottom part, we visualise relative prominence of the timescale contained in the spectrum of the system.
  The dominant timescale and its harmonics are highlighted in yellow.
  }\label{fig:spect_ENRON}
\end{figure}
Our change-point detection analysis of the ENRON dataset revealed an escalated frequency of activities during the unfolding of the ENRON scandal.
To delve deeper into this observation, we conduct a spectrum analysis of the temporal network dynamics, segmenting the data into two periods: before and after the year 2001.
The corresponding spectrograms, depicted in \cref{fig:spect_ENRON}, distinctly illustrate the shift in dominant timescales between these two eras.
Prior to the scandal, the prevailing timescale is notably slower compared to the period during the crisis.
This significant contraction in the dominant timescale post-2001 starkly illustrates the intensified pace of communication and internal dynamics as the company navigated through the scandal.
This analysis not only corroborates the findings from our change-point detection but also provides a quantitative measure of the organizational disruption during this period.

\paragraph{Empirical Data -- DEVS}

The DEVS dataset comprises 101 open-source software (OSS) projects, each subjected to a significant disruptive event--the departure of a core developer.
This `shock' led to varied outcomes: 64 projects failed to recover, with over half of these ceasing activity on GitHub within a year.
In contrast, 37 projects demonstrated their ability to recover from such a shock, continuing their activities up to the data cutoff point.

To shed light on the mechanisms underlying these divergent paths---failure versus recovery---we evaluate how the temporal spectrum of each project's dynamics varies over time.
Spanning from 100 weeks pre- to 50 weeks post-shock, we compute the timescales underlying the project dynamics for half-year periods.
To study their gradual evolution, we advance the period of analysis in one-week increments.
This approach allows us to extract the dominant timescales for 150 distinct periods: 74 pre-shock, 26 during the shock (serving as focus point for our analysis), and 50 post-shock.

\begin{figure}[ht]
  \centering
  \includegraphics[width=.6\textwidth]{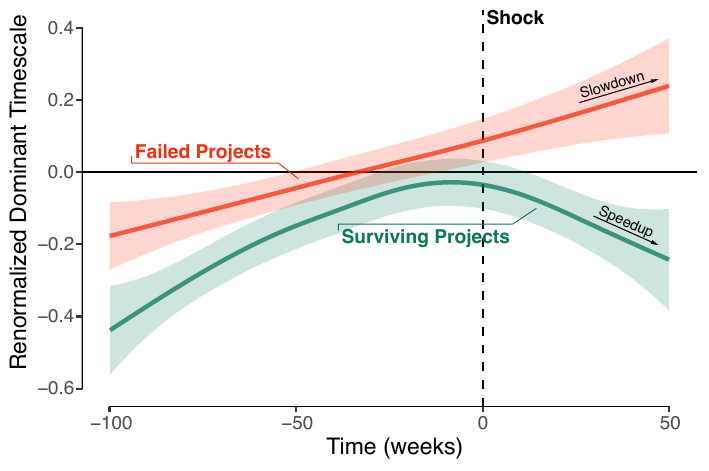}
  \caption{Temporal evolution of renormalized dominant timescales for projects' dynamics pre- and post-shock.
  This figure illustrates the evolution of the dominant timescales for 101 software projects over a period spanning from 100 weeks before to 50 weeks after a significant shock event.
  The timescales are renormalized by the standard deviation of each project's timescale across the observed timeframe to facilitate comparison.
  The average timescales obtained over periods overlapping the shock provide the reference point for the renormalisation.
  A generalized additive model (GAM) smoother with a cubic spline basis is applied to delineate trends separately for projects that failed (in red, 64/101) and those that survived (in green, 37/101) following the shock.
  Prior to the shock, all projects exhibit a notable increase in dominant timescale, indicating a universal slowdown.
  Post-shock, we observe a marked divergence in trend: surviving projects demonstrate a decrease in timescale, while failed projects continue to experience a slowdown.
  }\label{fig:spect_DEVS}
\end{figure}

\Cref{fig:spect_DEVS} showcases the evolution of these dominant timescales.
Distinguishing between projects that failed (depicted in red) and those that did not (in green), we visualise the trend followed by the dominant timescales with a generalized additive model smoother with a cubic spline basis.
Across all projects, we observe a slowdown before the shock--i.e., an increase in dominant timescales--indicative of decelerated project dynamics.
This commonality, however, diverges in the aftermath of the shock.

For projects unable to recover, the slowdown persists.
Conversely, our analysis reveals a dynamic reversal in surviving projects: a post-shock speedup.
This suggests a gradual return towards pre-shock operational conditions, emblematic of the projects' ability to respond to the shock.
Such a divergence in the characteristics between projects emphasizes the adaptive mechanisms within surviving projects, highlighting their capacity to overcome significant internal changes.

\section{Discussion}\label{sec:discussion}

Traditional methodologies in temporal network analysis have predominantly centered around identifying change-points, which mark distinct shifts in network dynamics.
While these methods are invaluable for detecting moments of transition, they tend to overemphasize these discrete points.
This obscures the actual timescales at which the network's underlying dynamics change.
Particularly in systems characterized by multiple different processes that overlap and intersect, focusing solely on change-points leads to a simplified interpretation that doesn't fully capture the complexity of the network's evolution.

Our work proposes a departure from standard methodologies that predominantly focus on identifying these discrete change-points.
By introducing a Bayesian modeling framework to infer and classify the relevant timescales in temporal networks, it enables a detailed exploration of temporal networks' dynamics, offering insights into the continuous evolution of these systems beyond mere moments of transition.
This perspective enriches our understanding of the complex processes driving temporal networks, enabling a holistic view of system dynamics and their broader implications.

Our analysis of the ENRON dataset reveals marked shifts in communication patterns, indicated by an acceleration in dominant timescales.
This suggests a speedup in the communication network's dynamics, mirroring the corporate crisis's escalation.
Similarly, our analysis of the GitHub developers' dataset demonstrates the effectiveness of our approach in discerning the impact of key events, such as a core developer's departure, on the projects' dynamics
Aligning with critical transition theory~\cite{scheffer2009early}, our findings highlight the importance of detecting early signs of change (e.g., slowdown) to prevent significant systemic shifts.
Our results underline the value of timescale analysis in identifying vulnerabilities early on, offering essential insights for the proactive management of complex systems and the formulation of strategies to mitigate potential crises effectively.

The case studies we have analyzed exhibited inherently different reactions to major perturbations.
The Enron Corporation succumbed to the aftermath of its accounting scandal, while many OSS projects analyzed displayed resilience to the loss of a core member of their projects.
At the heart of this lies a phenomenon pervasive in many complex systems: their adaptive nature~\cite{resilienceipp,schweitzer2021-fragileresilient,resilience_mega_paper}.
By providing a framework for analyzing the timescales governing the dynamics of adaptive systems, we can deepen our understanding of how systems can respond to and recover from disruptions, adapt to changes, and maintain their functionality over time.

\small \setlength{\bibsep}{1pt}

\normalsize
\section{Data and Methods}\label{sec:methods}

\subsection{Synthetic Data}\label{sec:methods:data:synth}
The model defined by \cref{eq:htcm_joint} is fully generative.
Exploiting this fact, in all the synthetic examples used throughout the article, we generated the data sampling from the model itself.
This serves as a perfect benchmark to showcase the features of our framework, as no additional assumptions beyond those already included in the model are needed to generate suitable data.

In the running example of \cref{fig:problem}, we generate the interaction data from the interplay of 2 simultaneous independent processes, highlighted in green and yellow in the figure.
Each process evolves at a different timescale: 22 timesteps for the green process and 33 timesteps for the yellow process.
To generate the data, we bind the output of two samples from the model, one with $\bf\Delta = (22,22,22,\dots)$, the second with $\bf\Delta = (33,33,33,\dots)$.
In both cases, we fix the number of node $N = 5$ and the expected number of edges per time-window $\hat m_\tau = 500$.
We plot the aggregated, static networks corresponding to each time-window in the top and bottom rows of \cref{fig:problem}.

When testing the resolution limits of the change-point detection method in \cref{sec:cpd:data}, we build the test data similarly.
The data is sampled from the model given a time-window partition $\bf\Delta_{GT} = (22,22,22,33,33,33,33,22,...)$, $N = 20$, and $T=105$.
To evaluate the resolution limits of the framework, we vary the total number of edges in the temporal network from 100 to 100,000.
Furthermore, to evaluate the impact of heterogeneity in the nodes, instead of sampling the degree sequences uniformly at random, we generate the data varying their coefficient of variance between 0 (all degrees the same) to 1.

\subsection{ENRON Dataset}
The Enron email data is a collection of email messages among employees of the Enron Corporation that has been published after the company filed for bankruptcy in 2001. 
In this article, we use the subset of the data as compiled by \citet{zhou2007strategies} and \citet{perry2013point}.
The data consists of 21,635 emails exchanged among 154 employees, resulting in a total of 38,388 timestamped edges spanning a period of over 3.6 years.

\subsection{DEVS Dataset}

The DEVS dataset consists of 101 open-source software (OSS) projects collected from GitHub.
\citet{russo2024shock} performed a rigorous selection process to identify, among the 125 million repositories, large OSS projects characterised by the departure of a core developer.
This process involved the following criteria:
(1) Exclusion of repositories with only one developer.
(2) Elimination of repositories with fewer than fifty commits.
(3) Removal of repositories with a duration between the oldest and newest commit of less than 100 days.
(4) Exclusion of repositories that were forked from pre-existing projects.
Furthermore, projects characterised by the departure of a core-developer were identified based on criteria such as the normalized skewness of commits distribution per developer, project duration, and the activity period of the core developer.
This curation resulted in a dataset comprising 101 major OSS projects, each characterized by a significant disruption event, involving a total of 8,234 active developers.

For each project, \emph{co-editing networks} were extracted using the Python library \texttt{gitnet}~\cite{gote2019git2net} to represent collaborations among developers based on the lines of code they edited together.
These co-editing networks are time-stamped, directed, multi-edge temporal networks, with each node representing a developer.
A directed edge from developer \(A\) to developer \(B\) indicates that \(B\) edited a line of code previously owned by \(A\).

\subsection{Temporal Configuration Models}\label{sec:methods:htcm}
\paragraph{A primer on Configuration Models}

Configuration models refers to random graph models preserving degree sequences (the configurations) exactly or in expectation under some uniformity assumptions~\cite{fosdick2018}.
In the context of this article, we consider only models that allow for multi-edges, i.e., for repeated edges incident to same pair of nodes.
These arise naturally in temporal networks as events between a pair of nodes can happen repeatedly over time, as can be seen in \cref{fig:problem}.

We can distinguish configuration models into two families.
(i) models for which the total number of edges is known a priori
and (ii) models for which the total number of edges is not known a priori but is generated by the model itself.
An example from the second family, is the Chung-Lu Configuration Model (CLCM)~\cite{Chung2002a,Norros2006}.
The Hypergeometric Configuration Model (HCM)~\cite{Casiraghi2021}, that we choose for this article, belongs to the first family, instead.
Its property, namely the fact the total number of edges is known a priori, is particularly suitable to model temporal networks.
Thanks to this property, modelling the distribution of the \emph{number} of edges over time can be separated from modelling the network \emph{structure}, simplifying the process.
In the next sections, we will show how this is achieved.
The formulation discussed in this article, nevertheless, can be extended to alternative models of either of these two families.

For simplicity, we present all models for the case of \emph{directed edges}.
Their undirected counterpart can be easily obtained from them.

\paragraph{Hypergeometric Configuration Model}\label{sec:hcm}

The hypergeometric configuration model (HCM)~\cite{Casiraghi2021,Casiraghi2017} fixes the total number of edges $m$ in a network $\G{}$ \emph{a priori}.
A network realisation is then generated by sampling these $m$ edges uniformly at random from an an urn containing all possible edges:
\begin{equation}\label{eq:hcm}
  \Pr(\G{}) = {\binom{\sum_{vw}\tout{v}\tin{w}}{m}}^{-1} \prod_{vw}{\binom{\tout{v}\tin{w}}{A_{vw}}}\;,
\end{equation}
where $\tout{v}\tin{w}\in\mathbb N_0$ is the total number of possible edges between a pair of nodes $v,w$ and $\tout v$ and $\tin v$ are parameters associated with each node $v$.
These parameters can be described as the node's \emph{activity} and \emph{attractiveness}, and can be related to the in- and out-degrees of the nodes through their expectations:
\begin{equation}\label{eq:exphcm1out}
  \mathbb E(\kout{v}) = \sum_w\frac{\tout{v}\tin{w}\cdot m}{\sum_{vw} \tout{v}\tin{w}}\;,
\end{equation}
\begin{equation}\label{eq:exphcm1in}
  \mathbb E(\kin{v}) = \sum_w\frac{\tout{w}\tin{v}\cdot m}{\sum_{vw} \tout{v}\tin{w}}\;.
\end{equation}
As the model's parameters appear always in pairs, they are defined modulo constants.
Thus, we are free to specify a constraint e.g., over their sum: $\sum_v\bm{\tout{}} = Q$ and $\sum_v\bm{\tin{}} = Q$.
Substituting back into \cref{eq:exphcm1out,eq:exphcm1in}, we get
\begin{equation}\label{eq:exphcm2out}
  \mathbb E(\kout{v}) = \sum_w\frac{\tout{v}\tin{w}\cdot m}{Q^2} = \tout{v}/Q\;,
\end{equation}
\begin{equation}\label{eq:exphcm2in}
  \mathbb E(\kin{v}) = \sum_w\frac{\tin{v}\tout{w}\cdot m}{Q^2} = \tin{v}/Q\;.
\end{equation}
Finally, choosing $Q=m$ ensures that (i) the total number of possible edges in the networks corresponds to the number of combinations of stubs or half-edges rewiring that gives $m$ directed edges~\cite{Casiraghi2021}, and (ii) the maximum likelihood estimator for $\bm{\tout{}}$ constrained by $\sum_v\bm{\tout{}} = m$ is $\bm{\kout{}}$.
If the parameters of the HCM are chosen to match their MLEs the model preserves degree sequences in expectation.

\paragraph{Hypergeometric Temporal Configuration Model} For the Hypergeometric Temporal Configuration Model (HTCM), the application of \cref{eq:tmodel} enables us to reformulate \cref{eq:hcm}.
We design independent urns for each time-window and uniformly distribute all multi-edges incident to each node pair among all time points within that window:
\begin{equation}\label{eq:HTCM_1}
  \Pr_\text{HTCM}(\bm{A_{vw\tau}}|A_{vw\tau},\Delta_\tau) = \frac{A_{vw\tau}!}{\prod_{t\in\tau}A_{vwt}!}\left(\frac1{\Delta_\tau}\right)^{A_{vw\tau}}
\end{equation}
\begin{equation}\label{eq:HTCM_2}
  \Pr_\text{HTCM}(\bm{A_\tau}|\bm{\tout{\tau}},\bm{\tin{\tau}}) = {\binom{\sum_{vw}\tout{v\tau}\tin{w\tau}}{m_\tau}}^{-1} \prod_{vw}{\binom{\tout{v\tau}\tin{w\tau}}{A_{vw\tau}}}\;,
\end{equation}
where \(\bm{A_{vw\tau}}\) represents the vector of all \(A_{vwt}\) for \(t\in\tau\), and \(\bm{A_\tau}\) denotes the adjacency matrix at time-window \(\tau\).

It is important to mention that \cref{eq:HTCM_1} is not the sole method for distributing multi-edges among time points within a window.
An alternative is the uniform probability \(p = \multiset{\Delta_\tau}{A_{ij\tau}}^{-1}\) across all configurations of \(A_{ij\tau}\) multi-edges over \(\Delta_\tau\) time points.
However, as our time-window definition assumes homogeneity within each window, and heterogeneity across windows, \cref{eq:HTCM_1} offers a more coherent representation.

Utilizing the fact that the \(\bm{\tout{\tau}},\bm{\tin{\tau}}\) are defined modulo constants for each time-window \(\tau\), we can specify constraints for each.
In the HTCM context, and its static counterpart, the constraints ensure that the total number of possible edges matches the number of possible in- and out-stubs combinations: \(\sum_{vw}\tout{v\tau}\tin{w\tau} = m_\tau^2\), yielding \(\sum_{v}\tout{v\tau}= \sum_{w}\tin{w\tau}=m_\tau\).
Combining this with \cref{eq:HTCM_1,eq:HTCM_2}, we obtain:
\begin{equation}\label{eq:HTCM_3}
  \Pr_\text{HTCM}(\bm{A_{\tau}}|\Delta_\tau,\bm{\tout{\tau}},\bm{\tin{\tau}}) = \frac1{\prod_{t\in\tau}A_{vwt}!}{\binom{m^2_\tau}{m_\tau}}^{-1} \prod_{vw}\frac{(\tout{v\tau}\tin{w\tau})!}{(\tout{v\tau}\tin{w\tau}-A_{vw\tau})!}\left(\frac1{\Delta_\tau}\right)^{A_{vw\tau}}\;.
\end{equation}

Assuming independence among time-windows, we arrive to the final equation of HTCM:
\begin{equation}\label{eq:HTCM_final}
  \Pr_\text{HTCM}(\G{\tau}|\bm\Delta,\bm{\tout{}},\bm{\tin{}}) = \frac1{\prod_{t}A_{vwt}!}\prod_{\tau}{\binom{m^2_\tau}{m_\tau}}^{-1} \prod_{vw}\frac{(\tout{v\tau}\tin{w\tau})!}{(\tout{v\tau}\tin{w\tau}-A_{vw\tau})!}\left(\frac1{\Delta_\tau}\right)^{A_{vw\tau}}\;.
\end{equation}

\subsection{A Generative Nonparametric Bayesian Model for temporal networks}\label{sec:methods:bayesian}

To balance model complexity---influenced by the number of time-windows $z$---and model fit---expressed by a model likelihood---we employ the Minimum Description Length (MDL) principle~\cite{rissanen1983}.
In the MDL framework, the optimal model is the one that \emph{minimizes} the total description length $\Sigma$
\begin{equation}\label{eq:DL}
\Sigma = \mathcal{S} + \mathcal{L},
\end{equation}
with \(\mathcal{S}\) denoting the number of bits needed to describe the temporal network given known model parameters, and \(\mathcal{L}\) representing the bits required to encode the model parameters.

From a Bayesian standpoint, the description length $\Sigma$ corresponds to the \emph{joint probability distribution} of the data $\G{T}$ and all discrete\footnote{Since the description length is a \emph{discrete code}, continuous model parameters cannot be encoded by it and needs to be integrated away.} model parameters $\bm\gamma$:
\begin{equation}
 \Sigma = - \log_2 \Pr(\G{T},\bm\gamma)\;.
\end{equation}
where the joint probability $\Pr(\G{T},\bm\gamma)$ is the full joint probability distribution $\Pr(\G{T},\bm\theta,\bm\Delta)$ integrated over all eventual continuous parameters.

The full joint distribution for the temporal configuration models is:
\begin{equation}\label{eq:joint}
\Pr(\G{T}, \bm{\tout{}},\bm{\tin{}}, \bm\Delta) = \Pr(\G{T}|\bm{\tout{}},\bm{\tin{}}, \bm\Delta) \cdot \Pr(\bm{\tout{}}| \bm\Delta) \cdot \Pr(\bm{\tin{}} | \bm\Delta) \cdot \Pr(\bm\Delta)\;.
\end{equation}
Here, the terms \(\Pr(\bm{\tout{}}| \bm\Delta)\), \(\Pr(\bm{\tin{}} | \bm\Delta)\), and \(\Pr(\bm\Delta)\) denote prior probabilities which we will specify subsequently.
This equation provides a complete generative model for both data and parameters.
Comparing \cref{eq:joint} with \cref{eq:DL}, we can easily see that \(\mathcal{S}\) corresponds to the likelihood of the model, and \(\mathcal{L}\) is given by the priors encoding the different model parameters.

If we write the posterior probability distribution for the model parameters \( \bm\gamma \)
\begin{equation}\label{eq:posterior}
\Pr(\bm\gamma | \G{T}) = \frac{\Pr(\G{T},\bm\gamma)}{\Pr(\G{T})},
\end{equation}
it becomes obvious that the parameter choices which \emph{minimise} the description length are those with the highest posterior probability.
The corresponding model is the most probable given the observed data.

In the sections that follows, we detail all required priors to express \cref{eq:joint}.
In doing so, we will take the approach outlined by~\citet{peixoto2015inferring} which have demonstrated its effectiveness in different network classification problems.

\paragraph{Prior for the Partition of Times into Time-Windows}

The first step in our Bayesian approach is to define a prior for the partition vector \(\bm\Delta\).
This vector specifies the width of each time-window and maps time points to specific time-windows.

A straightforward choice is to assign equal probability to all potential partitions:
\begin{equation}
  \Pr(\bm\Delta | z) = \abs{\Omega_z}^{-1}\,,
\end{equation}
where \(\Omega_z\) is the set of all valid partitions of an ordered sequence of \(T\) time points into \(z\) distinct time-windows.

The problem of partitioning a sequence of \(T\) elements into \(z\) bins echoes the classic \emph{stars and bars problem} in probability theory~\cite{pitman2012probability}.
Considering \(T\) elements, we have \(T-1\) gaps between them.
To distribute these elements across \(z\) non-empty bins, we insert \(z-1\) "separating bars" within these gaps.
The total number of possible configurations is given by:
\begin{equation}
  \abs{\Omega_z} = {\binom{T-1}{z-1}}\;.
\end{equation}

With this parametrization, we must still specify a \emph{hyper-prior} for the number of non-empty time-windows, \(z\).
Choosing again a uniform non-informative prior gives \(\Pr(z) = 1/T\), where \(z\) lies in the range [1, \(T\)].
This captures the notion that the number of time-windows can vary from a single giant window to \(T\) small windows, each containing one time point.
As this is essentially a multiplicative constant, it can be omitted in most inference scenarios.

Our nonparametric prior for the time-window partition is:
\begin{equation}\label{eq:priordeltamain}
  \Pr(\bm\Delta) = {\binom{T-1}{z-1}}^{-1}\frac1{T}\,.
\end{equation}

For the a fixed size time-window model, we must change the prior for the $\delta$ to:

\begin{equation}\label{eq:priordelta_fixedTWmain}
  \Pr(\bm\Delta) = \left(\frac{T(T-1)}{2}\right)^{-1}
\end{equation}

\paragraph{HTCM Prior for Node Activities}

As for the static configuration models, the parameters $\tout{v\tau}$ and $\tin{v\tau}$ encode the activity of node $v$.
A priori, a node can show any activity: from no activity, i.e., a weakly or strongly disconnected node, to maximum activity, i.e., all edges start or end from that node.
For the model we consider here, we choose non-informative priors for $\bm{\tout{}}$ and $\bm{\tin{}}$.

The number of interactions $M$ in the temporal network is known \emph{a priori}.
This means that we can model explicitly the number of interactions in each time-windows $m_\tau$.
Thus, given $m_\tau$, $\bm{\tout{}}$ and $\bm{\tin{}}$ represent the number of directed stubs (or half-edges) assigned to each node, as detailed in \cite{Casiraghi2021}.
Within each time-window containing $m_\tau$ multi-edges, the total number of stubs combinations $\sum_{vw}\tout{v\tau}\tin{v\tau}$ is equal to $m_\tau^2$.
A priori, we can assume that both parameters (that we denote as $\bm{\tstar{\tau}}$ for compactness) can take any combination of values in the interval $(0,m_\tau)$ that sums to $m_\tau$:
\begin{equation}\label{eq:priorthetahtcm}
  \Pr_\text{HTCM}(\bm{\tstar{\tau}}|m_\tau) = \multiset{N}{m_\tau}^{-1}\;,
\end{equation}
where $N$ is the number of nodes in the network, and $\multiset{N}{m_\tau} = \binom{N+m_\tau-1}{m_\tau}$ is the weak composition of $m_\tau$.

\Cref{eq:priorthetahtcm} assumes that all nodes of the network exist in all time-windows, but not necessarily all are active.
I.e., \cref{eq:priorthetahtcm} allows for some elements of $\bm{\tstar{\tau}}$ to be $0$.
If we had prior knowledge about the presence and absence of nodes in different windows, we could model the number of active node $n_\tau$ in a window separately.
However, we do not investigate this case here.

In order to remain nonparametric, we still require a prior for the vector of edges in each time-window $\bm m$.
In the temporal network $\G{T}$, there are in total $M$ edges distributed over $T$ times.
A priori, we can assume edges to be uniformly distributed among all times.
Therefore, we assume a uniform rate $M/T$ for every time.
The probability of having $m_\tau$ edges in a time-window of length $\Delta_\tau$, scales with $\Delta_\tau/T$.
Hence, the problem of distributing $M$ edges into $z$ time-windows corresponds to a multinomial process where $M$ is the number of trials, $z$ the number of mutually exclusive events (i.e., an edge is realised only into a specific time-window), and $p_\tau=\Delta_\tau/T$ are the event probabilities.
The prior for $\bm m$, conditioned on the lengths of each time-window, takes the form
\begin{equation}
  \Pr(\bm m | \bm\Delta, z) = \frac{M!}{\prod_{\tau=1}^{z}m_\tau!}\prod_{\tau=1}^{z}\left(\frac{\Delta_\tau}{T}\right)^{m_\tau}\;.
\end{equation}

\subsection{MCMC Inference Algorithm}\label{sec:mcmc}
To identify an optimal time-window partition $\bm\Delta$ we have to maximise its posterior distribution \cref{eq:posterior} or, equivalently, minimise the description length of the model.
To do so, we use a Markov Chain Monte Carlo importance sampling for the time-window partitions.

To generate a new time-window partition from an existing one, we randomly select one time-window to split into two new adjacent ones, two adjacent time-windows to join, or two adjacent time-windows whose separation point has to moved.
We then accept or reject the change such that after a long enough mixing time the time-window partitions are sampled according to the posterior distribution.

Looking at \cref{eq:htcm_joint}, it is evident that the equation can be easily factorised in different independent terms, each corresponding to the contribution of a given time-window.
Exploiting this fact, we can efficiently compute whether a change to the time-window partition improves the likelihood.
In more details, we proceed as follows.
Starting from an initial time-window partition $\bm\Delta^{(0)}$, we randomly choose whether to join two time-windows, move their separation point, or split one into two.
\begin{equation}
  \Pr(\tau_0 \to \{\tau_{1A},\tau_{1B}\}) = \Pr(\{\tau_{0A},\tau_{0B}\} \to \tau_1) = \Pr(\{\tau_{0A},\tau_{0B}\} \to \{\tau_{1A},\tau_{1B}\}) = \frac1 3\;.
\end{equation}

Let $\bm\Delta^{(0)}$ be the initial time-window partition and $\bm\Delta^{(1)}$ the new time-window partition obtained after performing one of the 3 changes.
If the new time-window partition yields a higher likelihood, we accept the change and proceed with a new one.
If the new time-window partition yields a lower likelihood, we accept the change with probability
\begin{equation}\label{eq:pchange}
  p_\text{change} = \exp\left[(\mathcal L_\text{HTCM}(\bm\Delta^{(1)}|\G{T}) + \mathcal L_\text{HTCM}(\bm\Delta_\text{Split}^{(0)}|\G{T}))/\beta\right]\cdot\frac{\Pr(\bm\Delta^{(1)}\to \bm\Delta^{(0)})}{\Pr(\bm\Delta^{(0)}\to \bm\Delta^{(1)})}\;,
\end{equation}
where $\beta$ is some \emph{temperature} and $\Pr(\bm\Delta^{(1)}\to \bm\Delta^{(0)})$ is the probability of performing the same change to revert from $\bm\Delta^{(1)}$ back to $\bm\Delta^{(0)}$.

\paragraph{Log-likelihood difference for splits and joins}
Let $\tau_0$ be a window to split, sampled uniformly at random among all $z$ time-windows in $\bm\Delta^{(0)}$.
We further sample a split point $t$ uniformly from the interval $[1,\Delta_\tau]$ of all times contained in the time-window $\tau$.
This generates two new time-windows $\tau_{1A}$ and $\tau_{1B}$ that are fully contained in the original window and replace it in the new time-window partition $\bm\Delta_\text{Split}^{(1)}$.

The difference in log-likelihood between the two time-window partitions for the HTCM can be easily obtained by computing $\mathcal L_\text{HTCM}(\bm\Delta_\text{Split}^{(1)}|\G{T}) - \mathcal L_\text{HTCM}(\bm\Delta^{(0)}|\G{T})$, noting that all terms not related to the modified window will cancel out.

In the case of joins, we proceed in a similar fashion.
Let $\{\tau_{0A},\tau_{0B}\}$ a pair of adjacent windows sampled uniformly at random from all the $z-1$ pairs of adjacent time-windows.
Joining the two windows result in a new time-window partition $\bm\Delta_\text{Join}^{(1)}$ with $z-1$ windows.
The differences in log-likelihood $\mathcal L_\star(\bm\Delta_\text{Join}^{(1)}|\G{T}) - \mathcal L_\star(\bm\Delta^{(0)}|\G{T})$ are given by the negatives of the split case. 
Finally, we can compute the probability of performing a specific join or a specific split and reverting these changes:
\begin{equation}
  \Pr(\bm\Delta^{(0)}\to \bm\Delta_\text{Split}^{(1)}) = \frac1z\frac1{\Delta_\tau-1}\;,
\end{equation}
\begin{equation}
  \Pr(\bm\Delta_\text{Split}^{(1)} \to \bm\Delta^{(0)}) = \frac1z\;,
\end{equation}
\begin{equation}
  \Pr(\bm\Delta^{(0)}\to \bm\Delta_\text{Join}^{(1)}) = \frac1{z-1}\;,
\end{equation}
\begin{equation}
  \Pr(\bm\Delta_\text{Join}^{(1)} \to \bm\Delta^{(0)}) = \frac1{z-1}\frac1{\Delta_\tau-1}\;,
\end{equation}
where $z$ is the number of time-windows $\abs{\bm{\Delta^{(0)}}}$ and $\tau$ is the time-window that will be split and has been joined, respectively.

\paragraph{Moves}
Moving the separation point between two adjacent time-windows $\tau_{0A}$ and $\tau_{0B}$ is achieved by performing a join and split consecutively, without checking the likelihood between the two operations.
The difference in log-likelihoods is similar to the case of splits with less terms, though, as the two partitions share the same number of time-windows $z$.
Moreover, in this case the probability of performing a move is identical to that of reverting it, as the same operations needs to be performed in the same order and the number of windows $z$ does not change.
Thus,
\begin{equation}
\frac{\Pr(\bm\Delta_\text{Move}^{(1)}\to\bm\Delta^{(0)})}{\Pr(\bm\Delta^{(0)}\to \bm\Delta_\text{Move}^{(1)})} = 1\;.
\end{equation}

\subsection{The Timescales Spectrum of \texorpdfstring{\(\G{T}\)}{\mathcal{G}(T)}}\label{sec:methods:thms}

Let's consider the description length (DL) expression introduced in the previous sections.
In the context of identifying the timescales of the processes underlying a temporal network \(\G{T}\), we can see this expression as a function of every timescale in the \([0,T]\) interval.
Specifically, while keeping the time-window size \(\Delta\) constant across time-windows, computing the joint probability as a function of \(\Delta\) provides us with a tool to evaluate how well different timescales fit the data.
Under some simplifying conditions, in fact, we can equate \emph{time-window size} with \emph{timescale}.
Assuming the data-generating process develops at a given timescale implies that the data should be partitioned into equally sized, adjacent time-windows of size equal the timescale, to match the process evolution.
We can formalise these principles as follows.
First, let's provide a formal definition of a \emph{change-point}.
\begin{definition}[Change-point]
	Let \(P\) be a dynamic data generating process characterised by two timescales: a fast timescale $t$ at which interactions are generated and a slow timescale $\tau$ at which the parameters governing how interactions are generated change.
	We define as \emph{change-point} the instant, usually measured in the timescale $t$ at which the data-generating process changes.
	If the slow timescale is constant over time, then the change-points for the process \(P\) are all distributed at an equal distance \(\tau\).
\end{definition}

\begin{thm}\label{thm:minima}
In a scenario with a single data-generating process at a constant timescale \(\tau=\Delta\), and under the assumption of a uniform prior on all timescales, negligible noise, and sufficient data resolution, the following holds true:
The time-window size \(\hat\Delta\) yielding the minimum description length (MDL) corresponds to the timescale \(\Delta\).
Thus, it perfectly aligns each time-window with each change point of the data generating process.
Local minima of the DL function are divisors (i.e., harmonics) of this MDL timescale (the fundamental frequency) or they are divisors of multiples of this fundamental frequency.
\end{thm}
\noindent In other words, local minima of the DL plot corresponds to either modelling timescales that identify \emph{perfectly} every change point in the data but possibly overfits it, by assuming a faster dynamics than the true one.
Alternatively, they correspond to timescales that underfit the data, but still manage to identify some of the existing change-points.
A proof of \cref{thm:minima} can be sketched as follows.

\textit{Part 1: Local Minima} \\
Consider a timescale \(\Delta_1\) that is able to identify at least one change point of the data generating process.
Given enough temporal resolution in the data, there exist a small \(\varepsilon>0\) such that any timescale \(\Delta'\), with \(\Delta_1-\varepsilon<\Delta'<\Delta_1+\varepsilon\), has approximately the same model complexity of \(\Delta_1\) but worse model fit, since it won't identify change points exactly.
This results in a higher DL as defined in \cref{eq:DL} compared to that of \(\Delta_1\).
Therefore, local minima of the DL function define timescales whose periods are divisors of multiples of the true timescale \(\Delta\).

\textit{Part 2: Divisors of \(\Delta\) (\(\Delta_d = \hat\Delta/d\))} \\
In the absence of noise, smaller windows within a window of length exactly \(\Delta\) define models with identical parameters.
In fact, no change points are present within such a window.
Therefore, while the data part of the description length in \cref{eq:DL} remains constant, the model complexity increases due to the presence of more time-windows, leading to a higher overall description length.

\textit{Part 3: Multiples of divisors of \(\Delta\) (\(\Delta_m = \Delta \cdot m / d\))} \\
Larger windows average over distinct change points, reducing data fit accuracy.
This results in a higher description length from the data part, which, in the presence of enough data points, cannot be offset by the reduced model complexity.
Therefore, among all local minima, \(\Delta\) itself has the smallest description length.

\bigskip

Under the ideal settings of \cref{thm:minima}, we can see that an analysis of the DL computed for all possible window sizes up to \(T\) provides us with an easy way to identify the true timescale of the data generating process---which corresponds to the window size that yields the minimum description length---as well as its harmonics---which correspond to the local minima of the DL function.
In more realistic scenarios, though, finite size effects originating e.g., from a coarse temporal resolution of the data, will complicate matters.
However, with enough data resolution we will always be able to identify the true timescale of the process.
Furthermore, we can extend \cref{thm:minima} to the case of multiple distinct processes.
In this case, however, local minima of the DL function will correspond to \emph{all} timescale matching change-points of one or more processes, complicating the issue.
\begin{thm}\label{thm:multiple}
In a scenario with \(n\) distinct process generating data at timescales \(\{\Delta_i\}_{i=1}^n\), and under the assumption of a uniform prior on all timescales, negligible noise, sufficient data, and high enough temporal resolution of the data, the following holds true:
The true timescales \(\{\Delta_i\}_{i=1}^n\) are harmonics of the MDL timescale \(\hat\Delta\).
\end{thm}
\noindent Again, we can easily see than the MDL \(\hat\Delta\) corresponds to the timescale that can match all change points.
However, this timescale does not necessarily correspond to the true timescale of any process, as it can be evinced from \cref{fig:problem}.
In fact, assuming absence of noise, enough data, and high enough data resolution, this is the greatest common divisor (GCD) among all true timescales.
Following the same reasoning of the proof sketched above, local minima will correspond to multiples or divisors of this optimal timescale.
Therefore, by the definition of GCD, the true timescales of the different processes underlying the system under study will appear as local minima in the DL plot.
The other local minima of the DL plot will further correspond to the harmonics of these true timescales and to their resonances, i.e., to combinations of these that match groups of change points.
We illustrate this in the top panel of \cref{fig:spectrogram}.

\bigskip

Theorems \ref{thm:minima} and \ref{thm:multiple} establish a critical understanding of local minima in the description length (DL) function as they relate to the timescales of processes in our temporal network \(\G{T}\).
However, every timescale that aligns with at least one change-point in the data generation processes can emerge as a local minimum.
This implies that merely cataloging these minima is not sufficient to detect the actual timescales of the underlying processes.
Addressing this, it becomes necessary to rank these identified timescales based on their significance within the context of the data generating process.

One initial method to assess the importance of timescales could be to directly examine the description length (DL) values of different local minima.
However, this approach has its limitations: 
as we move away from the minimum DL value, the DL inherently increases, leading to a biased favoring of timescales near the minimum description length (MDL) value.
To counter this limitation, we propose ranking timescales based on the relative prominence of their corresponding local minima.
We utilize the concept of \emph{topographic prominence}, applied to the maxima of the negative DL function, effectively transforming DL minima into peaks.

Topographic prominence, adapted from geomorphology, is a measure used to assess the relative height of a peak or valley within a landscape.
In our context, it quantifies the distinctiveness of each local maximum in the negative Description Length (DL) function.
The more a particular timescale (represented by a local maximum) stands out compared to its neighboring values, the higher its prominence.
This metric is useful in distinguishing between significant timescales that genuinely shape the data and those that are mere artifacts or less influential.
The prominence of a peak is determined by how much it rises above the lowest contour line that encircles it and no higher peak.

In our context, topographic prominence can be defined as follows.
Let \( \psi \) be a local maximum (peak) of the negative DL function.
The prominence of \( \psi \) is the vertical distance between \( \psi \) and the lowest contour line that surrounds \( \psi \) and no other higher peak.
Mathematically, if \( L \) is the negative DL of this lowest contour line, the prominence \( \text{Prom}(\psi) \) is given by:
\[
\text{Prom}(\psi) = -\text{DL}(\psi) + L
\]
This calculation emphasizes the relative importance of each timescale by measuring how much it stands out compared to its surroundings.

Independently of how well a timescale matches the overall data generating processes, we can expect that the true timescale of one of the underlying processes will have a larger improvement on its neighbourhood compared to its harmonics or to different resonances.
This means that the local minimum corresponding to the true timescale of one data generating process will be more prominent that that corresponding to one of its harmonics.

With this, we can propose a definition for the timescale spectrum of \(\G{T}\)'s dynamics.
\begin{definition}[Spectrum of a temporal network's dynamics]
	Let \(\psi\in[0,T]\) be a local minima of the DL function for a temporal network \(\G{T}\) and \(\text{Prom}(\psi)\) its prominence.
	Then, the spectrum of \(\G{T}\)'s dynamics is defined as the set of all \(\psi\in[0,T]\).
\end{definition}
Similarly to the spectrum of a signal in signal processing, the timescales spectrum of \(\G{T}\) contains the timescales needed to described the dynamics of the temporal network. 
Plotting the prominence of each local minimum of the DL function computed for some temporal network \(\G{T}\) provides us with the \emph{spectrogram} of \(\G{T}\)'s dynamics, highlighting the different timescales present in the data and their importance to the data generating process.
We showcase the spectrograms of the example introduced in \cref{fig:problem} in the bottom panel of \cref{fig:spectrogram}.

\end{document}